\documentstyle[epsfig]{elsart}

\def\PkE{P({\bf k},E)}
\def\PkkE{P({\bf k}_1,{\bf k}_2,E)}
\def\PkkEv{P_{var}({\bf k}_1,{\bf k}_2,E)}
\def\dPgr{\delta P_{GR}({\bf k}_1,{\bf k}_2,E)}
\def\dPint{\delta P_{INT}({\bf k}_1,{\bf k}_2,E)}

\begin{document}

\begin{frontmatter}

\title{
TWO-NUCLEON SPECTRAL FUNCTION IN INFINITE NUCLEAR MATTER
}

\author{Omar Benhar}
\address{INFN, Sezione Roma 1, I-00185 Rome, Italy}
\author{Adelchi Fabrocini}
\address{INFN and Dip.di Fisica, Universit\`a di Pisa, I-56100 Pisa, Italy}

\begin{abstract} 
 The two-nucleon spectral function in nuclear matter 
 is studied using Correlated Basis Function perturbation theory, 
 including central and tensor correlations produceded by a realistic 
 hamiltonian. 
 The factorization property of the two-nucleon momentum distribution 
 into the product of the two single nucleon distributions shows up 
 in an analogous property of the spectral function. The correlated 
 model yields a two-hole contribution quenched whith respect to 
 Fermi gas model, while the peaks acquire a quasiparticle width 
 that vanishes as the two momenta approach $k_F$. 
 In addition, three-hole one-particle and more complicated intermediate 
 states give rise to a background, spread out in energy 
 and absent in the uncorrelated models. The possible connections with 
 one- and two-nucleon emission processes are briefly discussed.
\end{abstract}

\begin{keyword}
Nuclear structure. Nuclear matter. Two-nucleon emission.
\end{keyword}

\end{frontmatter}

%%%%%%%%%%%%%%%%%%%%%%%%%%%%%%%%%%%%%%%%%%%%%%%%%%%%%%%%%%%%%%%%%%%%%%%%%%
\section{Introduction}
%%%%%%%%%%%%%%%%%%%%%%%%%%%%%%%%%%%%%%%%%%%%%%%%%%%%%%%%%%%%%%%%%%%%%%%%%%

The two-nucleon spectral function, yielding  the probability to remove 
two nucleons with momenta ${\bf k}_1$
 and ${\bf k}_2$ from nuclear matter leaving the 
residual system with excitation energy $E$, is defined as
(see, e.g., ref.\cite{Brown}):
\begin{equation}
\PkkE=\sum_n \vert \langle {\bar 0}|a^\dagger_{{\bf k}_1}a^\dagger_{{\bf k}_2}|
{\bar n}(A-2) \rangle \vert ^2
\delta (E-E_n(A-2)+E_0(A))\ ,
\label{TBSF}
\end{equation}
where $|\bar 0\rangle$ is the $A$-particle nuclear matter ground state with 
energy $E_0(A)$, $|\bar n (A-2)\rangle$ denotes a $(A-2)$-nucleon intermediate
 state with energy $E_n(A-2)$ and $a^\dagger_{{\bf k}}$ is
the usual creation operator. 

The two-nucleon spectral function carries relevant information on the 
short range structure of the nuclear medium and nucleon-nucleon (NN) 
correlations. Therefore, a considerable effort aimed at 
extracting experimental information on $\PkkE$ is currently being 
undertaken. In this context, a very important role is played by
electron-nucleus scattering experiments (a number of theoretical
and experimental topics in the field of electron-nucleus scattering
are reviewed in ref.\cite{book}).

Unambiguous evidence of strong correlation effects has been provided 
by single nucleon knock out $(e,e^\prime p)$ reactions, showing that in a 
nucleus only about 70~\% of the nucleons are in states of low momentum
and low removal energy, that can be described by a mean field theory.
The remaining 30 \% of the nucleons belong to strongly correlated pairs, 
whose occurrence is mainly to be ascribed to the one-pion-exchange tensor
force and to the repulsive core of the NN interaction.

The available $(e,e^\prime p)$ data at low missing momentum and low missing
energy clearly show the depletion of the occupation probabilities of single 
particle states predicted by the nuclear shell model (for a recent review see, 
e.g., ref.\cite{rmp}), thus providing a somewhat indirect 
measurement of correlation effects. Complementary information will soon be
available from single nucleon knock out experiments
specifically designed to investigate the kinematical region corresponding 
to large missing momentum and large missing energy, where correlation effects
are believed to be dominant\cite{TJNAFeep}.

Over the past decade, the availability of a new generation of high energy
 100 \% duty-cycle electron beams has made it possible to carry out 
double coincidence $(e,e^\prime NN)$ experiments, in which two knocked out
nucleons are detected. In principle, these experiments may allow for direct 
measurements of {\it correlation observables}, such as the momentum space
two-nucleon distribution or the two-nucleon spectral function. 
 However, as pointed out in ref.\cite{BFFee2N}, extracting 
the relevant dynamical information from the measured  $(e,e^\prime NN)$
cross section may turn out to be a challenging task, requiring both a 
careful choice of the kinematical setup and a quantitative theoretical
understanding of the final state interactions (FSI) of the knocked out 
nucleons.

Regardless of the difficulties involved in the interpretation of the
experimental data, reliable theoretical calculations of the two-nucleon
emission cross section carried out within 
the Plane Wave Impulse Approximation (PWIA), in which all the nuclear 
structure information is contained in the two-nucleon spectral function, 
have to be regarded as a minimal starting point, which can provide 
guidance for the optimization of future experiments. 

The basic assumptions underlying the PWIA scheme are that i) the exchanged 
virtual photon couples to either of the two outgoing nucleons and ii) FSI 
effects are negligible. The PWIA cross section of the process in which 
an electron of initial energy $E_i$ is scattered into the solid angle 
$\Omega_e$ with energy $E_f=E_i-\omega$,
while two nucleons of kinetic energies $T_p$ and $T_p^\prime$ are ejected into 
the solid angles $\Omega_p$ and $\Omega_p^\prime$, respectively, takes the
simple factorized form
\begin{equation}
\frac{d^9\sigma}{d\omega d\Omega_e d\Omega_p dT_p d\Omega_p^\prime dT_p^\prime}
= p(T_p+m)p^\prime(T_p^\prime+m) \sigma_{em} F_{{\bf p} {\bf p}^\prime}(p_m,E_m)\ ,
\label{PWIA:xsec}
\end{equation}
where $\sigma_{em}$ describes the structure of the electromagnetic vertex, 
$m$ is the nucleon mass and ${\bf p}$ and ${\bf p}^\prime$
denote the momenta of the detected nucleons. 
The missing momentum ${\bf p}_m$ and missing energy $E_m$ are defined as
\begin{equation}
{\bf p}_m = {\bf q} - {\bf p} - {\bf p}^\prime\ ,
\label{miss:mom}
\end{equation}
${\bf q}$ being the momentum transfer, and
\begin{equation}
E_m = \omega - T_p - T_p^\prime - T_R\ , 
\label{miss:en}
\end{equation}
where $T_R$ is the kinetic energy of the recoiling $(A-2)$-particle system.
The function $F_{{\bf p} {\bf p}^\prime}(p_m,E_m)$ appearing in 
eq.(\ref{PWIA:xsec}) can be written in terms of the two-nucleon spectral
functions $P({\bf p}-{\bf q},{\bf p}^\prime,E_m)$ and 
$P({\bf p},{\bf p}^\prime-{\bf q},E_m)$.

Pioneering studies of the PWIA two-nucleon emission cross section of the
$^{12}C(e,e^\prime p p)$ reaction are discussed in ref.\cite{Rosner}, 
whereas a G-matrix perturbation theory calculation of the $^{16}O$ two-proton
spectral function is described in ref.\cite{Muther}.
 
In this paper we discuss a calculation of the nuclear matter $\PkkE$ 
performed using Correlated Basis Function (CBF) perturbation theory. 
Our theoretical approach, in which the effects of the nonperturbative
components of the NN interaction are incorporated in the basis function 
using a variational approach, has proved to be particularly suited to describe 
quantities which are strongly 
affected by NN correlations, such as the single nucleon spectral function 
$\PkE$\cite{BFF}, the nucleon momentum distribution $n({\bf k})$\cite{Stefano_nk}
 and the off-diagonal density matrix 
$\rho({\bf r}_1,{\bf r}_1^\prime)$\cite{rho_1}.

Following ref.\cite{BFF}, we evaluate the dominant contributions to 
$\PkkE$ at the zero-th order of CBF, as well as the 
 perturbative corrections that have been shown to be relevant in 
the calculation of $\PkE$.

CBF theory of infinite nuclear matter is built on the set of 
{\em  correlated} states:
\begin{equation}
|N\rangle={\cal S}\left[\prod_{i<j} F(i,j)\right] |N\rangle_{FG}\ , 
\label{basis}
\end{equation}
obtained by applying a  symmetrized product of two-body 
correlation operators, $F(i,j)$, to the Fermi gas states
$|N\rangle_{FG}$.  The structure of $F(i,j)$ is similar to 
that of the NN interaction:
\begin{equation}
F(i,j)=\sum_n f^{(n)}(r_{ij}){\it O}^{(n)}(i,j)\ . 
\label{F:operator}
\end{equation}
The operators ${\it O}^{(n)}(i,j)$ include four central components ($1, 
({\bf\sigma}_i\cdot{\bf\sigma}_j), ({\bf\tau}_i\cdot{\bf\tau}_j), 
({\bf\sigma}_i\cdot{\bf\sigma}_j)({\bf\tau}_i\cdot{\bf\tau}_j$)) 
for $n=1,4$ 
and the isoscalar and isovector tensor ones $S_{ij}$ and 
$S_{ij}({\bf\tau}_i\cdot{\bf\tau}_j)$ for $n=5,6$. Additional spin-orbit 
components are sometimes introduced, but we will neglect them in this 
work. With transparent and widely employed notation, the components are 
also denoted as $c$ ($n=1$), $\sigma$, $\tau$ and $t$ (tensor).

Realistic correlated basis states are expected to be close to the 
eigenstates of the hamiltonian. An efficient recipe to choose the 
correlation 
functions consists in their variational determination 
by minimizing the expectation value of the hamiltonian in the 
correlated ground state: 
\begin{equation}
E_0^v=\langle 0|H|0\rangle\ .
\end{equation}
The scalar (or Jastrow) component $f^{(n=1)}(r)$ heals to unity 
for $r\rightarrow \infty$, whereas $f^{(n\neq 1)}(r)\rightarrow 0$. 
The tensor components have the longest range, since they are related to 
the one-pion exchange potential and  $F^{\dagger}(i,j)H_{ij}F(i,j)$ 
($H_{ij}$ is the hamiltonian associated with a two-nucleon pair) 
is a {\sl well-behaved} operator healing to $H_{ij}$ at large 
interparticle distances, like the G-matrix. The results presented 
in this work have been obtained using the Urbana $v_{14}$ potential 
supplemented by the TNI three-nucleon interaction \cite{U14}.

At the zero-th (or {\em variational}) order of CBF, $\PkkEv$ is  
obtained using the correlated states (\ref{basis}) in eq.(\ref{TBSF}). 
The variational estimate is then corrected by inserting perturbative 
corrections in the correlated basis. This procedure has been already 
adopted for many nuclear matter properties and it has been 
particularly successful in describing inclusive electromagnetic responses 
at both intermediate and high momentum transfers 
(see, e.g., ref.\cite{book}). 

In Section 2 we will discuss the relationships between the two-nucleon 
spectral function, the two-nucleon momentum distribution, 
 and the two-body density matrix, 
$\rho({\bf r}_1,{\bf r}_2;{\bf r}_1^\prime,{\bf r}_2^\prime)$.  
Section 3 will present theory and results for the variational 
two-nucleon  spectral function, as well as the main CBF perturbative 
corrections, evaluated with realistic interactions. Finally, Section 4
is devoted to conclusions and perspectives. 

%%%%%%%%%%%%%%%%%%%%%%%%%%%%%%%%%%%%%%%%%%%%%%%%%%%%%%%%%%%%%%%%%%%%%%%%%%%
\section{Spectral functions, momentum distributions and density matrices}
%%%%%%%%%%%%%%%%%%%%%%%%%%%%%%%%%%%%%%%%%%%%%%%%%%%%%%%%%%%%%%%%%%%%%%%%%%%

The nuclear matter two-body density matrix
$\rho({\bf r}_1,{\bf r}_2;{\bf r}_1^\prime,{\bf r}_2^\prime)$ is defined as
\begin{eqnarray}
\nonumber
\rho({\bf r}_1,{\bf r}_2;{\bf r}_1^\prime,{\bf r}_2^\prime)
 & = &  A(A-1) \\
 & \times  & \frac{ \int d^3r_3 \ldots d^3r_A\
\Psi_0^\dagger({\bf r}_1,{\bf r}_2,\ldots,{\bf r}_A)
\Psi_0({\bf r}_1^\prime,{\bf r}_2^\prime,\ldots,{\bf r}_A) }
{ \int d^3r_1 \ldots d^3r_A
|\Psi_0({\bf r}_1,{\bf r}_2,\ldots,{\bf r}_A)|^2}\ ,
\label{def:dm}
\end{eqnarray}
$\Psi_0({\bf r}_1,{\bf r}_2,\ldots,{\bf r}_A)$ being the ground state
wave function. For ${\bf r}_1={\bf r}_1^\prime$ and 
${\bf r}_2={\bf r}_2^\prime$, the density matrix reduces to the two-nucleon
density distribution $\rho({\bf r}_1,{\bf r}_2)$, yelding the joint probability 
of finding two nucleons at positions ${\bf r}_1$ and ${\bf r}_2$ in the 
nuclear matter ground state.

The two-nucleon momentum distribution, i.e. the probability that two nucleons 
carry momenta  ${\bf k}_1$ and ${\bf k}_2$, is related to the two-nucleon 
spectral function $\PkkE$, defined in  eq.(\ref{TBSF}), through
\begin{equation}
n({\bf k}_1,{\bf k}_2) = \int dE \PkkE\ =
\langle {\bar 0} | 
a^\dagger_{{\bf k}_1}a^\dagger_{{\bf k}_2}
a_{{\bf k}_2}a_{{\bf k}_1} | {\bar 0} 
\rangle\ .
\label{def:nkk}
\end{equation}

 From Eqs.(\ref{def:dm}) and (\ref{def:nkk}) it follows that 
$n({\bf k}_1,{\bf k}_2)$ can be written in terms of the density matrix 
according to 
\begin{equation}
n({\bf k}_1,{\bf k}_2) = \frac{1}{\nu^2}\frac{1}{\Omega^2}
\int d^3r_1 d^3r_2 d^3r_1^\prime d^3r_2^\prime
{\rm e}^{ i {\bf k}_1 \cdot {\bf r}_{11^\prime} } 
{\rm e}^{ i {\bf k}_2 \cdot {\bf r}_{22^\prime} }
\rho({\bf r}_1,{\bf r}_2;{\bf r}_1^\prime,{\bf r}_2^\prime)\ ,
\label{def2:nkk}
\end{equation}
where $\nu$ denotes the degeneracy of the system (in symmetric nuclear matter
$\nu$ = 4), $\Omega$ is the normalization volume,  
${\bf r}_{11^\prime}={\bf r}_{1} - {\bf r}_{1^\prime}$ and
${\bf r}_{22^\prime}={\bf r}_{2} - {\bf r}_{2^\prime}$. 

\begin{figure}
\begin{center}
\leavevmode
\centerline{\epsfig{figure=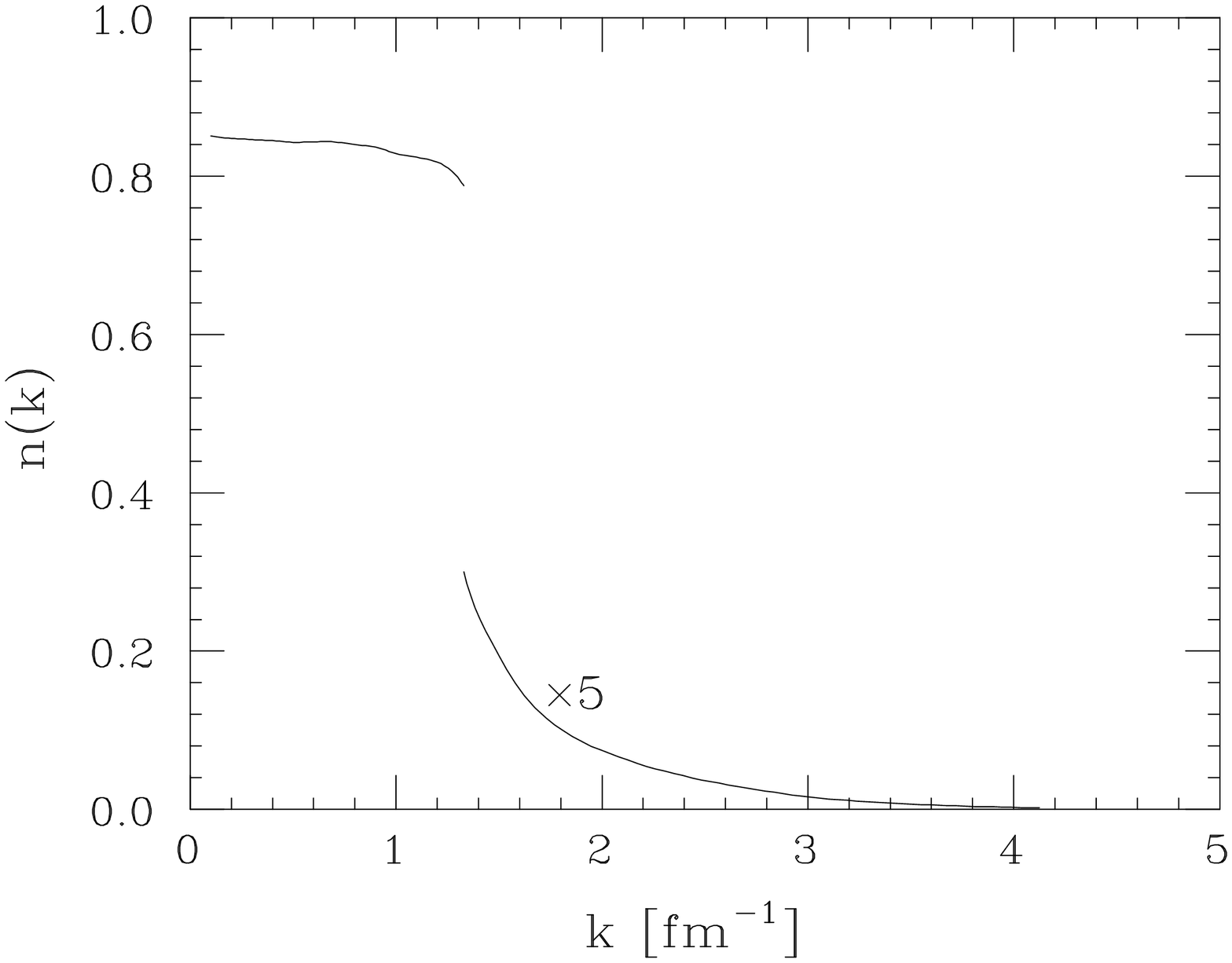,angle=090,width=17cm}}
\caption{
Single nucleon momentum distribution in nuclear matter, evaluated in
ref.{\protect\cite{Stefano_nk}} using CBF perturbation theory.
}
\label{nk}
\end{center}
\end{figure}

It is well known\cite{Stefano} that in a Fermi liquid {\em with no long
range order} the two-nucleon momentum distribution factorizes according to
\begin{equation}
n({\bf k}_1,{\bf k}_2) = n({\bf k}_1)n({\bf k}_2) + O \left( 
\frac{1}{A} \right)\ ,
\label{factor:nkk}
\end{equation}
where $n({\bf k})$ is the single nucleon momentum distribution, defined 
as
\begin{equation}
n({\bf k}) = \langle {\bar 0} |
a^\dagger_{{\bf k}}a_{{\bf k}} | {\bar 0} \rangle\ .
\label{def:nk}
\end{equation}
The $n({\bf k})$ resulting from the nuclear matter calculation of 
ref.\cite{Stefano_nk} is shown in fig. \ref{nk}. It exhibits a discontinuity
at $|{\bf k}| = k_F$ (the Fermi momentum $k_F$ is related to the density 
$\rho = A/\Omega$
through $k_F = (6 \pi^2 \rho/\nu)^{1/3}$), and a tail, extending to 
very large
momenta, reflecting the structure of the nuclear wave function at small 
interparticle distance. 
Eq.(\ref{factor:nkk}) implies that in the A $\rightarrow \infty$ limit  
$n({\bf k}_1,{\bf k}_2)$ contains the same dynamical information carried by 
$n({\bf k})$. 

Some insight on the role of short range correlations can be obtained rewriting
$n({\bf k}_1,{\bf k}_2)$ in terms of the relative and center of mass momenta
of the pair, defined as 
${\bf q} = ({\bf k}_1 - {\bf k}_2)/2$ and $Q = {\bf k}_1 + {\bf k}_2$, 
respectively, and studying the quantity
\begin{equation}
n_{rel}({\bf q}) = 4 \pi |{\bf q}|^2 \int d^3Q\ 
n\left( \left| \frac{{\bf Q}}{2} + {\bf q} \right| \right)
n\left( \left| \frac{{\bf Q}}{2} - {\bf q} \right| \right)\ .
\label{def:nq}
\end{equation}
Fig. 2 illustrates the behaviour of $n_{rel}({\bf q})$ at saturation density
(corresponding to $k_F = 1.33$ fm$^{-1}$), evaluated using 
the momentum distribution of ref.\cite{Stefano_nk} (solid line), compared to 
the prediction of the Fermi gas model (dashed line). 

The inclusion of 
NN interactions leads to a quenching of the peak, located at 
$|{\bf q}| \sim$ 0.7-0.8 fm$^{-1}$, and to the appearance of a sizeable
tail at $|{\bf q}| > k_F$. It is interesting to single out the contributions
to $n_{rel}({\bf q})$ corresponding to ${\bf k}_1, {\bf k}_2 < k_F$, 
${\bf k}_1< k_F$ and ${\bf k}_2 > k_F$, and ${\bf k}_1, {\bf k}_2 > k_F$.
They are represented in fig. 2 by diamonds, squares and crosses, respectively
(note that the results shown by the crosses have been enhanced by a factor
 10). The ${\bf k}_1, {\bf k}_2 < k_F$ component provides about 72 $\%$ of the 
normalization of $n_{rel}({\bf q})$ and becomes vanishingly small at 
$|{\bf q}|>k_F$. On the other hand, the contributions coming from pairs
in which at least one of the nucleon momenta is larger that $k_F$ 
are much smaller in size (the curves marked with squares and crosses yield 
26 $\%$ and 2 $\%$ of the normalization, respectively) 
but extend up to $|{\bf q}| \sim$ 4 fm$^{-1}$. Similar results  
have been recently obtained \cite{Arturo} using the momentum distribution resulting from 
a G-matrix perturbation theory calculation \cite{Ramos}.

%%%%%%%%%%%%%%%%%%%%%%%%%%%%%%%%%%%%%%%%%%%%%%%%%%%%%%%%%%%%%%%%%%
\section{CBF calculation of the two-nucleon spectral function}
%%%%%%%%%%%%%%%%%%%%%%%%%%%%%%%%%%%%%%%%%%%%%%%%%%%%%%%%%%%%%%%%%%

The propagation of two hole states in nuclear matter is described 
by the hole-hole part of the two-body Green's function\cite{Brown}, 
\begin{equation}
G_{hh}({\bf k}_1,{\bf k}_2, E)=
\langle \bar 0 \vert a^{\dagger}_{{\bf k}_1}a^{\dagger}_{{\bf k}_2}
\frac{1}{H-E_0-E-\imath\eta}
a_{{\bf k}_2} a_{{\bf k}_1} \vert \bar 0 \rangle\ ,
\label{2body:green}
\end{equation}
where $H$ is the nuclear matter hamiltonian. $G_{hh}$ is defined for 
$ \epsilon(=-E)<2e_F$, $e_F$ being the Fermi energy. 

The two-nucleon spectral function, given in eq.(\ref{TBSF}), is 
straightforwardly related to the imaginary part of $G_{hh}$ by
\begin{equation}
\PkkE= {1 \over {\pi}} \Im G_{hh}({\bf k}_1,{\bf k}_2, E)\ .
\label{TBSF_1}
\end{equation}

CBF perturbation theory employes the set of correlated states 
given in eq.(\ref{basis}). Note that the basis states are non orthogonal
to each other. Orthogonalization can been implemented using 
 the procedure proposed in ref.\cite{ortho}, 
which preserves the diagonal matrix elements of the hamiltonian between 
correlated states. 
This technique has been succesfully applied to the calculation of the 
one-body spectral function in ref.\cite{BFF} 
(hereafter denoted I). 

The perturbative expansion is obtained 
separating the nuclear hamiltonian $H$ into two parts: 
$H=H_0+H_I$, and using standard Rayleigh-Schr\"odinger type expansions.
The unperturbed and interaction terms, denoted by $H_0$ and $H_I$, respectively, 
are defined through
\begin{equation}
\langle N \vert H_0 \vert M \rangle = 
\delta_{NM} \langle N \vert H_0 \vert N \rangle
=\delta_{NM} E_N^v\ ,
\end{equation}
\begin{equation}
\langle N \vert H_I \vert M \rangle = 
(1-\delta_{NM}) \langle N\vert H_I\vert M \rangle\ .
\end{equation}

The expectation value in eq.(\ref{2body:green}) is calculated by 
expanding 
\begin{equation}
\vert {\bar 0}\rangle = {
{\sum_n(-)^n\left[\left(H_0-E_0^v\right)^{-1}
\left(H_I-\Delta E_0\right)\right]^n\vert 0 \rangle}
\over
{\vert\langle {\bar 0}\vert{\bar 0}\rangle \vert^{1/2}} }\ ,
\label{CBF2}
\end{equation}
where $\Delta E_0=E_0-E_0^v$. Perturbative corrections to the
intermediate states $\vert N\rangle$ are taken into account 
carrying out an analogous expansion in powers of 
$(H_I-\Delta E_0)$ for the propagator appearing in eq.(\ref{2body:green}), 
as discussed in I: 

\begin{eqnarray}
\nonumber
\frac{1}{(H-E_0-E-i\eta)} & = & \frac{1}{(H-E_0^v-E-i\eta)} \\
                          & \times & \sum_n(-)^n 
\left\{ \left(H_I-\Delta E_0\right) \frac{1}{(H-E_0^v-E-i\eta)} 
\right\}^n\ .
\end{eqnarray}

We will now describe the structure of the $n=0$ (i.e. variational) and higher 
order terms of the CBF perturbative expansion 
of $\PkkE$.

\subsection{The variational spectral function}

At the lowest order in the perturbative expansion the
spectral function is given by
\begin{equation}
\PkkEv=\sum_n \vert \langle  0|a^\dagger_{{\bf k}_1}
a^\dagger_{{\bf k}_2}|  n (A-2) \rangle \vert ^2 
\delta (E-E^v_n(A-2)+E^v_0(A))\ ,
\label{TBSF_v}
\end{equation}
where the correlated states (\ref{basis}), and their energies 
$E^v_0(A)$ and  $E^v_n(A-2)$ are used.

The main contributions to the sum in eq.(\ref{TBSF_v}) come 
from correlated two-hole ($2h$) and 
three-hole one-particle ($3h1p$) intermediate states. In the 
uncorrelated Fermi gas only $2h$ states contribute to 
$P({\bf k}_1,{\bf k}_2,E)$, leading to the well known result
\begin{eqnarray}
\nonumber
P_{2h,FG}({\bf k}_1,{\bf k}_2,E) & = &
\Theta (k_F-|{\bf k}_1|) \Theta (k_F-|{\bf k}_2|) \\
 & \times & \delta \left( E + 
\frac{\hbar ^2}{2m}|{\bf k}_1|^2 + \frac{\hbar ^2}{2m}|{\bf k}_2|^2
\right)\ , 
\label{P2h_FG}
\end {eqnarray}
where $\Theta(x)$ denotes the usual step function. The presence of 
correlations, besides changing the structure of the $2h$ contribution 
in the way we will discuss below, allows for nonvanishing contributions 
from other intermediate states, with a consequent quenching of 
the $2h$ peak since part of the strength is moved to higher excitation 
energies. 
 
The variational $2h$ spectral function is given by:
\begin{eqnarray}
\nonumber
P_{2h,v}({\bf k}_1,{\bf k}_2,E) & = &
  {1 \over {2}} \sum_{{\bf h}_1,{\bf h}_2}
\vert \Phi_{{\bf k}_1,{\bf k}_2}^{{\bf h}_1,{\bf h}_2} \vert^2
\Theta (k_F-|{\bf k}_1|) \Theta (k_F-|{\bf k}_2|) \\
& \times & \delta(e^v_{h_1}+e^v_{h_2}+ E)\ , 
\label{P2h_v}
\end{eqnarray}
where $e^v_h=\langle {\bf h}\vert H \vert {\bf h}\rangle  - E_0^v$ is 
the variational single particle energy and the $2h$ overlap matrix 
element, $\Phi_{{\bf k}_1,{\bf k}_2}^{{\bf h}_1,{\bf h}_2}$, is 
\begin{equation}
\Phi_{{\bf k}_1,{\bf k}_2}^{{\bf h}_1,{\bf h}_2}=
   \langle 0 \vert a^{\dagger}_{{\bf k}_1}  a^{\dagger}_{{\bf k}_2}  
   \vert {\bf h}_1,{\bf h}_2\rangle\ .
\label{Phi_2h}
\end{equation}

The correlated $1h$ overlap matrix element,  $\Phi_{\bf k}^{\bf h} 
  =\langle 0 \vert a^{\dagger}_{{\bf k}}  \vert {\bf h}\rangle$, was 
computed in I using cluster expansion techniques. The same method is used 
for the $2h$ overlap, with the result that only {\em unlinked} 
cluster diagrams ({\em i.e.} diagrams where the points reached by the 
$k_{1,2}$-lines are not connected to each other by any dynamical, 
$f^2-1$ and $f-1$, or statistical correlations) contribute. 
The {\em linked} diagrams contribution 
turn out to be of order 1/A, and therefore vanish in nfinite nuclear 
matter. Note that the factorization property of the 
two-nucleon momentum distribution, illustrated by eq.(\ref{factor:nkk}),  
is also a consequence of the fact that {\em linked} 
diagrams do not contribute in the A$\rightarrow \infty$ limit.
The definitions, as well as the technicalities 
entering the cluster expansion method can be found in I and in the 
related references and will not be repeated here. 
Because of the factorization property, in nuclear matter we obtain 
\begin{equation}
\Phi_{{\bf k}_1,{\bf k}_2}^{{\bf h}_1,{\bf h}_2}=
\Phi_{{\bf k}_1}^{{\bf h}_1}\Phi_{{\bf k}_2}^{{\bf h}_2}
\delta_{{\bf h}_1-{\bf k}_1}\delta_{{\bf h}_2-{\bf k}_2}\ ,
\label{Phi1_2h}
\end{equation}
the CBF expression of $\Phi_{\bf k}^{\bf h}$ being given in I 
(eq.(A.19)). 

The correlated $2h$ spectral function retains the Fermi gas delta 
shaped peak. However, correlations quench the peak itself (via the 
$2h$ overlap matrix element) and move it to $E=-e^v_{h_1}-e^v_{h_2}$. 

The $3h1p$ contribution reads
\begin{eqnarray}
\nonumber
P_{3h1p,v}({\bf k}_1,{\bf k}_2,E) & = &  \frac{1}{6} 
\sum_{{\bf h}_1,{\bf h}_2,{\bf h}_3,{\bf p}_1}
\vert \Phi_{{\bf k}_1,{\bf k}_2}
^{{\bf h}_1,{\bf h}_2,{\bf h}_3,{\bf p}_1} 
\vert^2 \\
& \times & \delta(e^v_{h_1}+e^v_{h_2}+e^v_{h_3}-e^v_{p_1} + E)\ , 
\label{P3h1p_v}
\end{eqnarray}
%\end{equation}
the overlap matrix element being given by
\begin{equation}
\Phi_{{\bf k}_1,{\bf k}_2}
^{{\bf h}_1,{\bf h}_2,{\bf h}_3,{\bf p}_1}= 
   \langle 0 \vert a^{\dagger}_{{\bf k}_1}  a^{\dagger}_{{\bf k}_2}  
   \vert {\bf h}_1,{\bf h}_2,{\bf h}_3,{\bf p}_1 \rangle\ .
\label{Phi_3h1p}
\end{equation}

In the limit A$\rightarrow \infty$ the contributions of fully linked diagrams 
vanish in the $3h1p$ overlap as well. As a consequence, 
\begin{equation}
\Phi_{{\bf k}_1>k_F,{\bf k}_2>k_F}
^{{\bf h}_1,{\bf h}_2,{\bf h}_3,{\bf p}_1} = 0\ . 
\label{Phi1_3h1p}
\end{equation}
In the other cases we obtain
\begin{equation}
\Phi_{{\bf k}_1<k_F,{\bf k}_2>k_F}
^{{\bf h}_1,{\bf h}_2,{\bf h}_3,{\bf p}_1}
=\sum
\Phi_{{\bf k}_1}^{{\bf h}_1}
\Phi_{{\bf k}_2}^{{\bf h}_2,{\bf h}_3,{\bf p}_1}
\delta_{{\bf h}_1-{\bf k}_1}\ ,
\label{Phi2_3h1p}
\end{equation}
and
\begin{equation}
\Phi_{{\bf k}_1<k_F,{\bf k}_2<k_F}
^{{\bf h}_1,{\bf h}_2,{\bf h}_3,{\bf p}_1}
=\sum \left[
\Phi_{{\bf k}_1}^{{\bf h}_1}
\Phi_{{\bf k}_2}^{{\bf h}_2,{\bf h}_3,{\bf p}_1}
\delta_{{\bf h}_1-{\bf k}_1} + 
{\bf k}_1 \leftrightarrow {\bf k}_2 \right]\ , 
\label{Phi3_3h1p}
\end{equation}
where the sums include all permutations of the ${\bf h}_i$ 
indices, while the structure of the 
$2h1p$ overlap $\Phi_{{\bf k}}^{{\bf h}_1,{\bf h}_2,{\bf p}_1}=  
   \langle 0 \vert a^{\dagger}_{{\bf k}}   
   \vert {\bf h}_1,{\bf h}_2,{\bf p}_1 \rangle$  
is discussed in I (Appendix B).

Unlike the $2h$ contribution, $P_{3h1p,v}$ is no longer delta shaped, 
but it is rather spread out in energy. 
It may be interpreted as a background contribution to be 
added to the two-quasiparticle part of the spectral function.

The next intermediate state to be considered is the $4h2p$ one, 
whose contribution may be written in terms of products of two 
squared  $2h1p$ overlap matrix elements. However, it is numerically 
negligible with respect to the $3h1p$ term (several order of magnitudes 
smaller) and can be safely neglected.

The above intermediate states completely exhaust the momentum 
distribution sum rule in nuclear matter at the variational level:
\begin{eqnarray}
\nonumber
 n^{v}({\bf k}_1,{\bf k}_2) & = &
 {\frac {\langle  0\vert 
a^{\dagger}_{{\bf k}_1} a^{\dagger}_{{\bf k}_2} 
a_{{\bf k}_2}a_{{\bf k}_1}\vert  0 \rangle }
 {\langle  0\vert 0 \rangle }} =
 n^{v}({\bf k}_1)n^{v}({\bf k}_2)\\
&=&\int dE \PkkEv\ .\\
\nonumber
\label{SR_var}
\end{eqnarray}
 
\subsection{Perturbative corrections}

The admixture of 
$m$-hole $n$-particle correlated states in $\vert \bar N \rangle$ 
originate CBF perturbative corrections. As far as the ground state is 
concerned, they correspond to  the $n\geq 1$ terms in eq.(\ref{CBF2}).  

Following the strategy developped in I, the perturbative corrections are 
classified as 
$\dPgr$ and $\dPint$, coming from ground and intermediate state corrections, 
respectively. 
The total spectral function is then given by
\begin{equation}
\PkkE=\PkkEv+\dPgr+\dPint\ .
\end{equation}

Here we consider $2h2p$ admixtures to the ground state and  $2h1p$  
admixtures to the $1h$ intermediate states. The first ones contribute 
for all $|{\bf k}|$ values (both below and above $k_F$), and their 
contribution 
can be expressed in terms of the correlation self energy 
$\Sigma^{CO}({\bf k},E)$, whose imaginary part is given by
\begin{equation}
\Im\Sigma^{CO}({\bf k},E)={\frac{\pi}{2}}\sum \vert 
\langle 0\vert H \vert {\bf k}{\bf p}_2{\bf h}_1{\bf h}_2\rangle \vert^2
\delta (E+e^v_{p_2}-e^v_{h_1}-e^v_{h_2})\ ,
\label{Corr1}
\end{equation}
at $|{\bf k}|>k_F$ and $E<e^v_F$ and 
\begin{equation}
\Im\Sigma^{CO}({\bf k},E)={\frac{\pi}{2}}\sum \vert 
\langle 0\vert H \vert {\bf p}_1{\bf p}_2 {\bf k}{\bf h}_2 \rangle \vert^2
\delta (E+e^v_{h_2}-e^v_{p_1}-e^v_{p_2})\ ,
\label{Corr2}
\end{equation}
at $|{\bf k}|<k_F$ and $E>e^v_F$.
The self energy is computed using correlated states and retaining 
 two- and three-body separable contributions in the 
 cluster expansion of the relevant matrix element \cite{fantoni82}.

The $2h1p$ intermediate state contribution involves 
the polarization  self energy $\Sigma^{PO}({\bf k},E)$. The corresponding
imaginary part reads
\begin{equation}
\Im\Sigma^{PO}({\bf k},E)={\frac{\pi}{2}}\sum \vert 
\langle {\bf k} \vert H \vert{\bf p}_2{\bf h}_1{\bf h}_2\rangle\vert^2
\delta (E+e^v_{p_2}-e^v_{h_1}-e^v_{h_2})\ ,
\label{Polar1}
\end{equation}
at $|{\bf k}|<k_F$ and $E<e^v_F$ and
\begin{equation}
\Im\Sigma^{PO}({\bf k},E)={\frac{\pi}{2}}\sum \vert 
\langle {\bf k} \vert H \vert{\bf p}_1{\bf p}_2 {\bf h}_2\rangle\vert^2
\delta (E+e^v_{h_2}-e^v_{p_1}-e^v_{p_2})\ ,
\label{Polar2}
\end{equation}
at $|{\bf k}|>k_F$ and $E>e^v_F$. 

The effects of these corrections on the spectral functions at 
$|{\bf k}_1|,|{\bf k}_2|<k_F$ are:
{\em i)} a quenching of the variational two-quasiparticle strength 
and a shift of its position by a quantity 
$\delta e_{k_1}^{CO}+\delta e_{k_2}^{CO}(k)$,  
with
\begin{equation}
\delta e_k^{CO}={\frac {1}{\pi}} \int_{e^v_F}
^{\infty} dE\ {\frac {\Im \Sigma^{CO} ({\bf k},E)}
{e^v_k-E}}\ , 
\label{deltaecoh} 
\end{equation}
from ground state corrections, and {\em ii)} a broadening of the 
delta-like peak itself, roughly proportional to 
$\Im \Sigma ^{PO}({{\bf k}_1},E=-e^v_{k_1}) + 
\Im \Sigma ^{PO}({{\bf k}_2},E=-e^v_{k_2})$, from intermediate state 
admixtures. 

The expression for the total $\delta P$ at 
$|{\bf k}_1|,|{\bf k}_2|<k_F$ reads 
\begin{eqnarray}
\nonumber
\delta P({\bf k}_1,{\bf k}_2,E) & = & 
-P_{2h,v}({\bf k}_1,{\bf k}_2,E)\\
\nonumber
& + &{1\over{\pi}}\Im \left\{
\left[ \vert \Phi_{{\bf k}_1,{\bf k}_2}^{{\bf k}_1,{\bf k}_2} \vert^2 +
\Sigma_2^{PO}({\bf k}_1,E)+\Sigma_2^{PO}({\bf k}_2,E)\right]\right. \\
\nonumber
& \times &\left[ \alpha({\bf k}_1,{\bf k}_2)(-e^v_{k_1}-e^v_{k_2}-E)
 -\delta e_{k_1}-\delta e_{k_2} \right. \\
\nonumber
& - & \Re \left( 
\Sigma^{PO}({\bf k}_1,E)- \Sigma^{PO}({\bf k}_1,E=-e^v_{k_1}) + 
  1\rightarrow 2 \right) \\
& - & i \left. \left. \left(
\Sigma^{PO}({\bf k}_1,E) + 1\rightarrow 2 \right) \right]^{-1} \right\}\ ,
\label{deltaP2} 
\end{eqnarray}
where $\alpha({\bf k}_1,{\bf k}_2)=1-\delta n_2({\bf k}_1)-
\delta n_2({\bf k}_2) -\delta n^\prime_2({\bf k}_1)-
\delta n^\prime_2({\bf k}_2)$. The quantities $\delta n_2({\bf k})$, 
$\delta n^\prime_2({\bf k})$ and $\Sigma_2^{PO}({\bf k},E)$ are defined 
in I.

Fig. \ref{spec:1} gives two examples of the two-nucleon spectral function 
below $k_F$. 
The figure shows the variational one-hole ($P_{1h,v}$) and 
two-hole ($P_{2h,v}$) peaks and the total spectral function 
(solid line). The one-nucleon spectral function below $k_F$ can be 
separated into a {\em single particle} and a {\em correlated} part
\cite{benhar94}. The former originates from $1h$ intermediate states 
and their admixtures, the latter from $n$-hole $(n-1)$-particle states, 
and it would be strictly zero in absence of correlations. 
The corresponding 
contributions for the two-nucleon spectral functions are shown in the 
figure as dashed and dot-dash lines, respectively. The correlated 
part has been enhanced by a factor 10$^2$, while the single particle 
and total ones by a factor 10. The figure also gives the barely 
visible variational correlated part (dotted lines).

Ground state $2h2p$ corrections give the leading contributions 
to $\delta P$ when one of the momenta is above $k_F$. So, for 
$|{\bf k}_1|<k_F$ and  $|{\bf k}_2|>k_F$, we obtain 
\begin{eqnarray}
\nonumber
\delta P({\bf k}_1,{\bf k}_2,E) & = & {1\over{\pi}} \Im 
\left\{ \vert\Phi_{{\bf k}_1}^{{\bf k}_1} \vert^2  
{\Sigma^{CO}({\bf k}_2,E+e_{k_1})\over
{(E+e_{k_1}+e_{k_2})^2}} \right.\\
& + & \left.
\Phi_{{\bf k}_1}^{{\bf k}_1}  
{\Sigma_4^{GR}({\bf k}_2,E+e_{k_1})\over
{E+e_{k_1}+e_{k_2}}}\right\}\ ,
\label{deltaP22} 
\end{eqnarray}
where $\Sigma_4^{GR}({\bf k},E)$ is again given in I.

The spectral functions for two such cases are shown in fig. \ref{spec:2}, 
together with the one-hole variational peaks corresponding to the momenta 
below $k_F$. The whole spectral function now comes from the correlated 
part. The figure gives the total (solid lines) and variational (dotted 
lines) $P$, as well as the perturbative corrections (dashed lines).

Corrections at $|{\bf k}_1|, |{\bf k}_2| > k_F$
 are quadratic in the self energies and 
have not been considered.

\section{Summary and conclusions}

We have carried out a calculation of the nuclear matter two-nucleon 
spectral function $\PkkE$ within the framework of CBF perturbation theory. 
The zero-th order approximation, corresponding 
to a variational estimate, has been supplemented with higher order 
corrections associated with both $2h2p$ admixture to the ground 
state and  $2h1p$ admixtures to the $1h$ states. 

The results show that the inclusion of NN interactions produces 
drastic changes in the behavior of $\PkkE$, with respect to the 
predictions of the Fermi gas model. The peaks correponding to the 
single particle states get quenched already at zero-th order, and
aquire a width when higher order corrections are included.  
Due to strong short range NN correlations, the strength removed from
the quasi particle peak is pushed to large values of momenta and removal
energy, giving rise to a broad background.  

The factorization property exhibited by the two-nucleon momentum 
distribution of a normal Fermi liquid has been found to hold at the
level of the amplitudes entering the calculation of $\PkkE$ as well.
As a result, all the linked cluster contributions to the two-nucleon
spectral function vanish in the infinite nuclear matter limit. 

In principle, the results discussed in this paper may be used to estimate
{\em correlation observables}, related to the small components of $\PkkE$, 
whose measurement represents the ultimate goal of the $(e,e^\prime NN)$ 
experimental programs. However, it has to be emphasized that, in order 
to make fully quantitative 
predictions, our approach should be extended to finite systems, working
along the line proposed in refs.\cite{nuclei,rho_1}. The analysis 
of the A dependence of $\PkkE$ would also provide very interesting 
information on the relevance of the 1/A linked contributions and the 
applicability of the factorized approximations for the two-nucleon
spectral function and momentum distribution. 

\begin{ack}

The authors are grateful to S. Fantoni, G.I. Lykasov, A. Polls and
W.H. Dickhoff for many illuminating discussions.

\end{ack}

%%%%%%%%%%%%%%%%%%%%%%%%%%%%%%%%%%%%%%%%%%%%%%%%%%%%%%%%%%%%%%%%%%%%%%%%%

\clearpage

%%%%%%%%%%%%%%%%%%%%%%%%%%%%%%%%%%%%%%%%%%%%%%%%%%%%%%%%%%%%%%%%%%%%%%%%%
%%%%%%%%%%%%%%%%%%%%%%%%%%%%%%%%%%%%%%%%%%%%%%%%%%%%%%%%%%%%%%%%%%%%%%%%%%

\clearpage

%%%%%%%%%%%%%%%%%%%%%%%%%%%%%%%%%%%%%%%%%%%%%%%%%%%%%%%%%%%%%%%%%%%%%%%%%
\begin{figure}
\begin{center}
\leavevmode
\centerline{\epsfig{figure=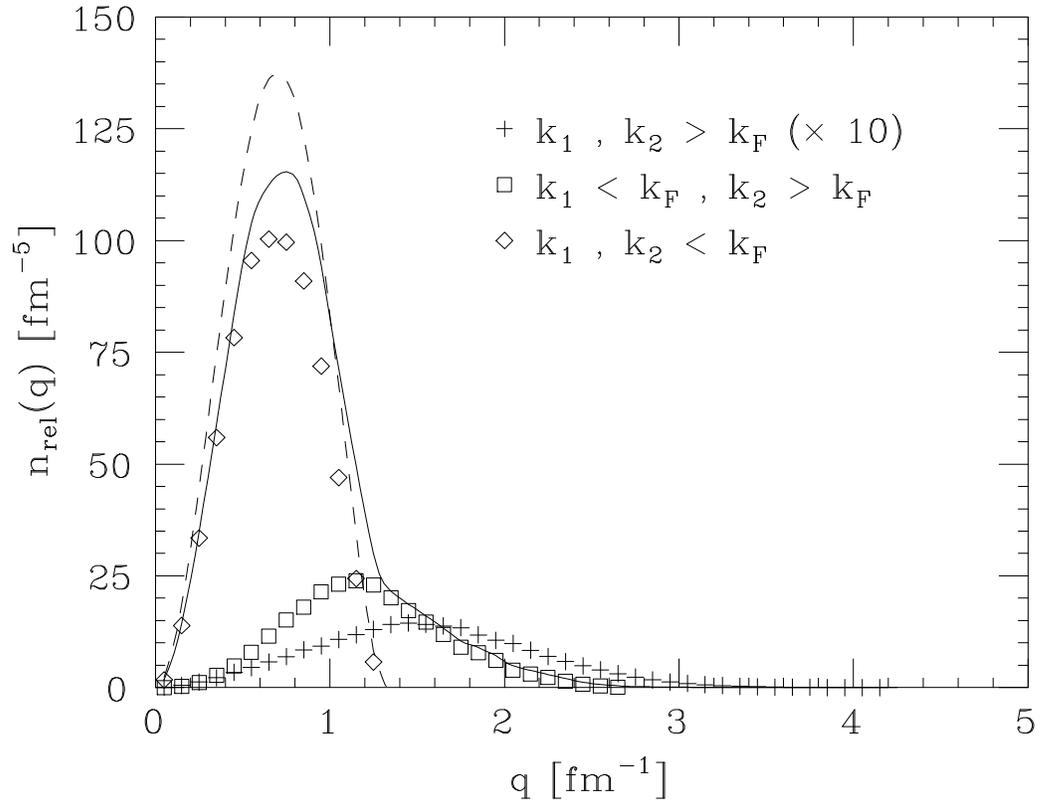,angle=090,width=17cm}}
\caption{
The distribution $n_{rel}({\bf q})$ defined in eq.(\protect\ref{def:nq}).
The solid and dashed lines represent the results obtained using
CBF perturbation theory and the Fermi gas model, respectively.
The meaning of the other curves is explained in the text.
}
\label{nq}
\end{center}
\end{figure}
%%%%%%%%%%%%%%%%%%%%%%%%%%%%%%%%%%%%%%%%%%%%%%%%%%%%%%%%%%%%%%%%%%%%%%%%%

\clearpage

%%%%%%%%%%%%%%%%%%%%%%%%%%%%%%%%%%%%%%%%%%%%%%%%%%%%%%%%%%%%%%%%%%%%%%%%%
\begin{figure}
\begin{center}
\leavevmode
\centerline{\epsfig{figure=fig3_np.epsi,angle=000,width=8cm}}
\caption{
Two examples of two-nucleon spectral function at
$|{\bf k}_1|, |{\bf k}_2| < k_F$. The different curves are discussed
in the text.
}
\label{spec:1}
\end{center}
\end{figure}
%%%%%%%%%%%%%%%%%%%%%%%%%%%%%%%%%%%%%%%%%%%%%%%%%%%%%%%%%%%%%%%%%%%%%%%%%

\clearpage

%%%%%%%%%%%%%%%%%%%%%%%%%%%%%%%%%%%%%%%%%%%%%%%%%%%%%%%%%%%%%%%%%%%%%%%%%
\begin{figure}
\begin{center}
\leavevmode
\centerline{\epsfig{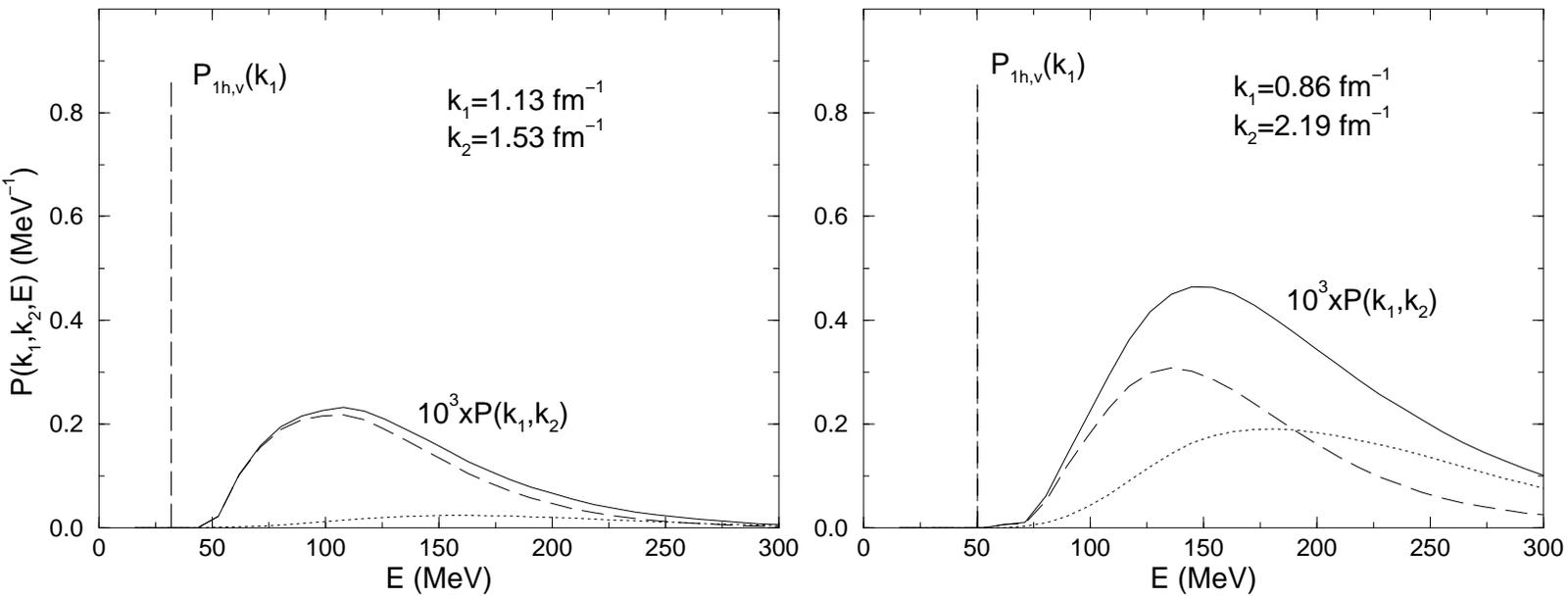}}
\caption{
Two examples of two-nucleon spectral function at
$|{\bf k}_1| < k_F, |{\bf k}_2| > k_F$. The different curves are
discussed in the text.
}
\label{spec:2}
\end{center}
\end{figure}
%%%%%%%%%%%%%%%%%%%%%%%%%%%%%%%%%%%%%%%%%%%%%%%%%%%%%%%%%%%%%%%%%%%%%%%%%

\end{document}